\documentclass[sn-nature,iicol]{sn-jnl}

\usepackage{graphicx}%
\usepackage{multirow}%
\usepackage{amsmath,amssymb,amsfonts}%
\usepackage{amsthm}%
\usepackage{mathrsfs}%
\usepackage[title]{appendix}%
\usepackage{xcolor}%
\usepackage{textcomp}%
\usepackage{manyfoot}%
\usepackage{booktabs}%
\usepackage{algorithm}%
\usepackage{algorithmicx}%
\usepackage{algpseudocode}%
\usepackage{listings}%

\usepackage{graphicx}%
\usepackage{multirow}%
\usepackage{amsmath,amssymb,amsfonts}%
\usepackage{amsthm}%
\usepackage{mathrsfs}%
\usepackage[title]{appendix}%
\usepackage{xcolor}%
\usepackage{textcomp}%
\usepackage{manyfoot}%
\usepackage{booktabs}%
\usepackage{algorithm}%
\usepackage{algorithmicx}%
\usepackage{algpseudocode}%
\usepackage{listings}%
\usepackage{microtype}
\usepackage{mathtools}
\usepackage[braket, qm]{qcircuit}
\usepackage{physics}
\usepackage{graphicx}
\usepackage{bm}
\usepackage[caption=false]{subfig} 
\usepackage{comment}
\usepackage{tabularx}
\usepackage{xcolor}
\usepackage{nicefrac}

\newcommand{\xdata}{\mathbf{x}}
\newcommand{\xdataset}{X}
\newcommand{\ydataset}{y}

\begin{document}

\title[Reinforcement learning-based architecture search for quantum machine learning]{Reinforcement learning-based architecture search for quantum machine learning}

\author[1,2]{\fnm{Frederic} \sur{Rapp}}\email{frederic.rapp@ipa.fraunhofer.de}

\author[1]{\fnm{David A.} \sur{Kreplin}}

\author[1,2]{\fnm{Marco F.}\sur{Huber}}

\author*[1]{\fnm{Marco} \sur{Roth}}\email{marco.roth@ipa.fraunhofer.de}

\affil[1]{\orgname{Fraunhofer Institute for Manufacturing Engineering and Automation IPA}, \orgaddress{\street{Nobelstrasse 12}, \city{Stuttgart}, \postcode{70569}, \country{Germany}}}

\affil[2]{\orgdiv{Institute of Industrial Manufacturing and Management IFF}, \orgname{University of Stuttgart}, \orgaddress{\street{Allmandring 35}, \city{Stuttgart}, \postcode{70569}, \country{Germany}}}

\abstract{
Quantum machine learning models use encoding circuits to map data into a quantum Hilbert space. While it is well known that the architecture of these circuits significantly influences core properties of the resulting model, they are often chosen heuristically. In this work, we present a novel approach using reinforcement learning techniques to generate problem-specific encoding circuits to improve the performance of quantum machine learning models. 
By specifically using a model-based reinforcement learning algorithm, we reduce the number of necessary circuit evaluations during the search, providing a sample-efficient framework.
In contrast to previous search algorithms, our method uses a layered circuit structure that significantly reduces the search space. Additionally, our approach can account for multiple objectives such as solution quality, hardware restrictions and circuit depth. We benchmark our tailored circuits against various reference models, including models with problem-agnostic circuits and classical models. Our results highlight the effectiveness of problem-specific encoding circuits in enhancing QML model performance.
}


\maketitle
\section{Introduction}

Quantum machine learning (QML) is a promising application of noisy intermediate-scale quantum computing (NISQ)~\cite{Preskill_2018}, and has gained significant attention in recent years~\cite{cerezo2022qml}. Central to many QML approaches is the use of encoding circuits, which map data into a quantum Hilbert space. These circuits may optionally be parameterized and are crucial components of methods such as quantum neural networks (QNN)~\cite{mitarai2018, Benedetti_2019} and quantum kernel methods~\cite{Schuld_2019, Havl_ek_2019, Huang_2021}.

Unlike variational algorithms used for quantum simulation~\cite{TILLY20221, Peruzzo2014} or optimization~\cite{farhi2014quantum}, there are few restrictions when it comes to the design and structure of quantum circuits used for QML. They can be tailored to accommodate hardware-specific constraints such as connectivity or native gates.

This design freedom allows for the creation of circuits that are well-suited to specific hardware and problem constraints. However, if the encoding circuit is not chosen properly, optimization can become infeasible~\cite{ragone2023unified, larocca2024review} and poorly chosen circuits can result in inadequate models~\cite{kuebler2021inductive}. While there are some established principles such as using data re-uploading and hardware efficient gate sets, they may not necessarily maximize the model's performance. From the no free lunch theorem~\cite{585893} it is well known that optimal performance is achieved by a problem specific model. Although this has been confirmed for QML models~\cite{kuebler2021inductive, bowles2023contextuality}, there are no established guidelines for how to construct QML encoding circuits that optimize the performance of the resulting model for a given task.

As a remedy, recent work has focused on incorporating properties of the data, such as symmetries, into QML architectures to provide models with a informed inductive bias~\cite{ragone2023representation, PRXQuantum3030341, PRXQuantum4010328, Schatzki2022tfq}. Nonetheless, many of these approaches require expert knowledge of problem domains and quantum circuit design, a prerequisite shared by other informed design choices~\cite{gili2023inductive}. This gap motivates the development of automated methods that can create and evaluate different encoding circuits, capable of generating problem-specific encoding circuits from input data.

\begin{figure*}[t]
    \includegraphics[width=\textwidth]{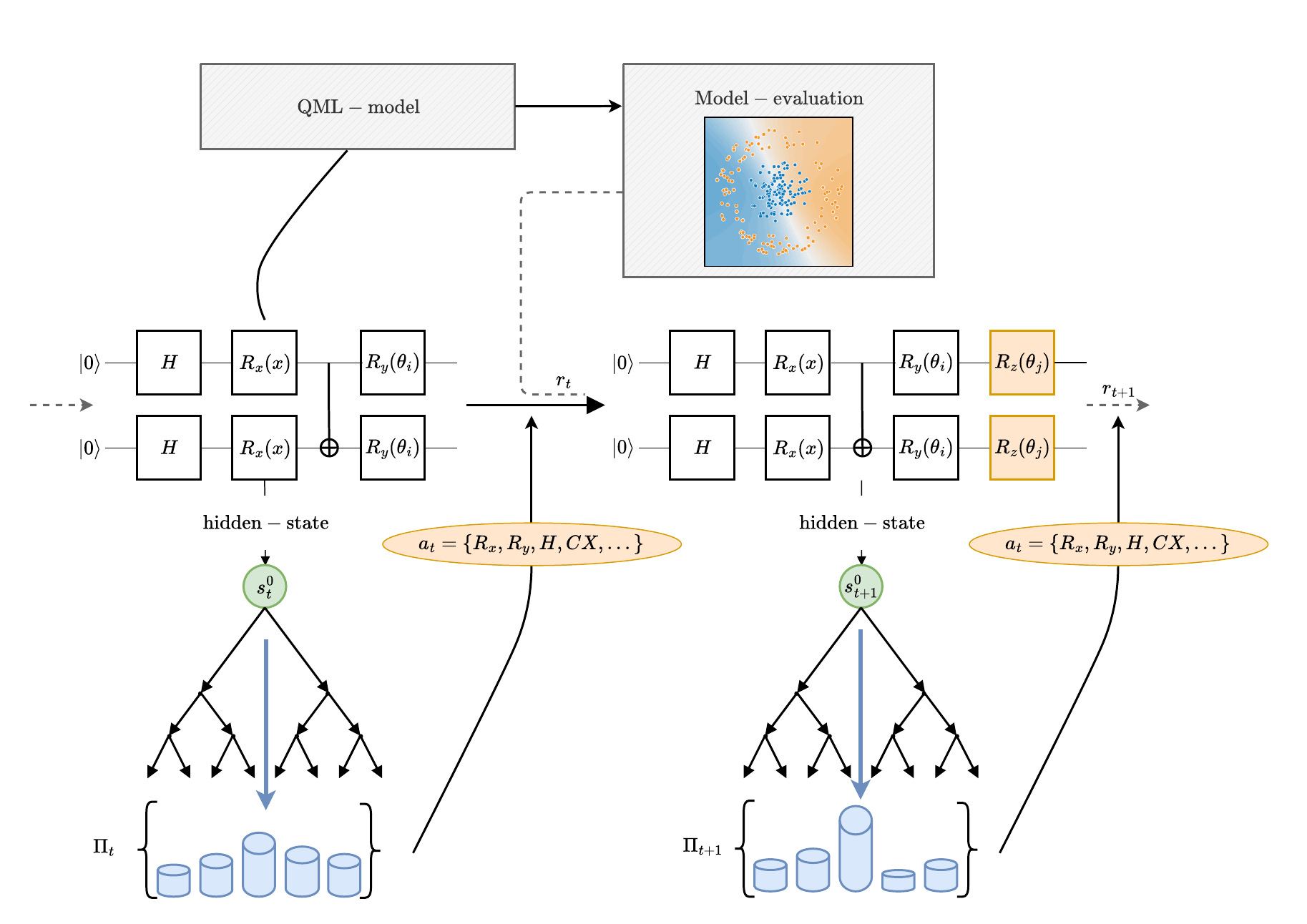}
    \caption{Automatic encoding circuit generation sketch. The MuZero algorithm observes the current circuit configuration, and transforms it into a hidden state representation $s_t^k$. By performing a Monte Carlo tree search for each time step $t$ the model obtains a policy $\Pi_t$ from which an action $a_t$ is sampled \cite{Schrittwieser_2020}. The action corresponds to choosing a unitary operator (e.g., $H, CX,...$) from a set of operators $\mathcal{V}$ that is appended on the current encoding circuit in a layered fashion (the same gate for each qubit). Each circuit $U_t$ is used to encode data for a QML model. Based on the cross-validation score of this model, the MuZero agent receives a reward $r_t$ at each time step. The calculation of the immediate rewards, value function, prediction function and policy during the Monte Carlo tree search, and the replay buffer handling follows the concept described in~\cite{Schrittwieser_2020}, using fully-connected neural networks to approximate the functions.
    }
    \label{fig: concept_layout}
\end{figure*}

In this work, we introduce a novel approach for QML circuit generation using model-based reinforcement learning (RL). In RL, an agent develops a policy by interacting with an environment to maximize expected cumulative rewards. For circuit generation, the observations correspond the current circuits, actions determine the next layer of quantum gates, and rewards are based on the QML model's training performance measured by the cross-validation score. While this setup could be realized with several RL algorithms, in this work we focus on MuZero~\cite{Schrittwieser_2020}. The algorithm uses deep neural networks to map the interaction between circuit architecture and QML model performance into a hidden state representation, significantly reducing the need for detailed environmental information. Actions are selected in a Monte Carlo tree search~\cite{Swiechowski2023} fashion, which relies on internal actions and rewards, reducing the number of evaluations of the QML model compared to model-free approaches. Figure~\ref{fig: concept_layout} illustrates our method's conceptual layout. 

Using a quantum support vector machine (QSVM)~\cite{Rebentrost_2014} based on a projected quantum kernel (PQK)~\cite{Huang_2021}, we demonstrate that our \emph{MuZero circuit search} (MCS) algorithm can generate problem-specific encoding circuits that surpass both reference circuits from the literature and those derived from random sampling or genetic search algorithms~\cite{Altares_Lopez_2021}. The results are consistently favorable across three different data sets: two involving regression tasks and one classification task. The resulting QML models perform comparably or slightly superior to various classical ML models.

The remainder of this work is structured as follows. In Sec.~\ref{sec: priorwork}, we provide an overview of prior work in the field of quantum circuit architecture search. Section~\ref{sec: methods} defines the general problem setting and explains the MCS algorithm. In Sec.~\ref{sec: results}, we demonstrate the versatility and circuit-creating capabilities of our algorithm by benchmarking it on several data sets. We conclude with a discussion in Sec.~\ref{sec:discussion}.

\subsection{Prior work}
\label{sec: priorwork}

The integration of RL techniques into quantum circuit design has been explored in various approaches. While these studies utilize RL for quantum circuit architecture search, their primary applications lie outside of QML. Additionally, in contrast to our work, they rely on model-free RL approaches. 

One key aspect focuses on optimizing quantum circuits to meet quantum hardware requirements, such as reducing gate count or enhancing qubit routing efficiency~\cite{Pozzi.2022}. Another aspect is the use of neural architecture search for general quantum circuit design~\cite{Zhang_2022,Zhang_2021}. A notable contribution in the area of general quantum circuit optimization is the methodology proposed in~\cite{ruiz2024quantum}, which employs tensor networks to reduce the number of gates in quantum circuits. This framework focuses on minimizing the gate overhead in deep circuit architectures, thereby facilitating practical implementations on quantum hardware.
In~\cite{NEURIPS2021_97244127} the authors apply RL techniques to create and optimize quantum circuits for variational quantum eigensolvers, aiming to enhance accuracy while maintaining a low circuit depth.
The work of~\cite{altmann2024challenges} addresses the challenges associated with using RL for quantum circuit design, introducing the concept of using Markov decision processes to guide the application of continuously parameterized quantum gates to optimize quantum circuits, specifically in the context of quantum state preparation.

In the context of QML the work of~\cite{incudini2023automatic} proposes an architecture search for quantum kernels using techniques from neural architecture search and AutoML. Among other techniques, such as Bayesian optimization, they also employ RL optimization based on a model-free state–action–reward–state–action algorithm. While~\cite{incudini2023automatic} aims to construct novel quantum kernels using combinatorial optimization techniques, our work focuses on the data-driven construction of encoding circuits for fixed QML models and quantum kernels to maximize the performance of the chosen method. The work of~\cite{Altares_Lopez_2021} introduces a multi-objective genetic algorithm for the automatic generation of quantum feature maps for fidelity-based quantum kernels. We include a slightly modified version of their implementation in our benchmarks in Sec.~\ref{sec: benchmark}.

Notably, this work goes beyond these methods. In contrast to previous work, we focus on a data-driven approach to optimize the performance of a specific QML model using a model-based RL algorithm to reduce the necessary number of circuit evaluations. Our layered circuit approach effectively reduces the search space. Additionally, we benchmark our method on various tasks, including regression and classification problems which demonstrates that our approach not only matches but often surpasses the performance of previously proposed genetic algorithms.

\section{Methods}
\label{sec: methods}

Consider a model trained on a data set $\mathcal{D}=(\xdataset,\ydataset)$, with $\xdataset=\lbrace \xdata_1,\dots,\xdata_N\rbrace$ comprising $N$ data points $\xdata_i\in\mathcal{X}\subset\mathcal{R}^{d}$ with corresponding labels $y_i\in\mathcal{Y}$. We are interested in regression problems with $\mathcal{Y}\subset{\mathbb{R}}$, and binary classification tasks with $\mathcal{Y}=\lbrace 0,1 \rbrace$. 
Quantum machine learning models rely on embedding data  $\xdata\in\mathcal{X}$ into quantum states
\begin{equation}
    \rho(\xdata,\pmb{\theta}) = U(\xdata, \pmb{\theta}) \rho_0 U(\xdata, \pmb{\theta})^\dagger \,,
\end{equation}
with $\rho_0 = \dyad{0}$, and parameters $\pmb{\theta} = (\theta_1, \theta_2, \ldots, \theta_N)^\top, \quad \text{where } \theta_i \in [-2\pi, 2\pi] \text{ for } i = 1, 2, \ldots, N$. The operator $U(\xdata, \pmb{\theta})$ is typically realized by a quantum circuit, and can usually be expressed as a product of unitary operators
\begin{equation}
    U(\xdata,\pmb{\theta})=\prod_{l=1} V_l(\xdata,\pmb{\theta})\,,   
\label{eq:encoding_circuits}
\end{equation}
where $V_l(\xdata,\pmb{\theta})=\exp(-i\phi_l(\xdata,\pmb{\theta})H_l)$. Here $H_l$ are traceless Hermitian operators and $\phi_l$ is a (potentially) non-linear function of the parameters and the data. Note that this includes the special cases $\phi_l=x^j$ and $\phi_l=\theta^j$, where $x^j$ and $\theta^j$ are the $j$-th element of a datapoint $\xdata$ and parameter vector $\pmb{\theta}$, respectively. 

A QML model can then be obtained by training the parameters $\pmb{\theta}$ and  calculating the expectation value with respect to some observable $O$ as $f(\xdata,\pmb{\theta}) = \tr\left(\rho(\xdata,\pmb{\theta})O\right)$. 
This is usually done by minimizing an appropriate loss function. The resulting optimized circuit can be used to derive various quantum models, such as QNNs or quantum kernel methods like QSVMs.

The choice of encoding circuit plays a defining role in the model's architecture. To account for the current hardware restrictions in the NISQ regime, the set of operators $\mathcal{V}$ from which the operators $V_l$ in Eq.~\eqref{eq:encoding_circuits} are chosen is often restricted to the native gate set of a specific hardware, resulting in hardware-efficient circuits that require a minimal amount of transpilation. 
 
The circuit architecture, i.e., the choice of operators and their placement in the encoding circuit  profoundly impacts the properties of the resulting model. For example, the expressivity of QML models is tightly connected to the circuit structure~\cite{Schuld_2021}. Furthermore, certain encoding circuits in quantum kernel methods can cause the quantum kernel to resemble a classical kernel, losing potential quantum advantage~\cite{Schuld_sup_a_kernels}. Using an appropriate circuit is therefore crucially important to achieve a well performing QML model. 

Despite its importance, this aspect is often neglected. A notable exception where informed architecture design is used is geometric QML. Here, symmetries in the data are reflected in the design of the circuit, resulting in models with desirable properties such as good generalization and trainability~\cite{ragone2023representation, PRXQuantum3030341, PRXQuantum4010328, Schatzki2022tfq, PRXQuantum5020328}. However, tailoring the encoding requires deep knowledge about the underlying properties of the data set, which is often intractable. For general data sets and ML problems, the choice of circuit architecture is therefore an open question which is usually not addressed beyond heuristics. 

In this work, we develop an RL algorithm to address this problem. The algorithm creates encoding circuits iteratively by choosing and placing unitary operators such that the performance of a resulting QML model is optimized. The individual components of the framework will be discussed in the following, an overview of the algorithm can be found in Fig.~\ref{fig: concept_layout}.

\subsection{Model-based reinforcement learning algorithm}

An RL agent interacts with its environment by taking actions $a_t\in \mathcal{A}$ at discrete time steps $t$ and receiving a reward $r_t\in\mathcal{R}$ in return. Here, $\mathcal{A}$ and $\mathcal{R}$ are discrete sets of actions and rewards, respectively. The agent's objective is to optimize its policy $\Pi_t:\mathcal{S}\rightarrow\mathcal{A}$ in order to maximize cumulative rewards, where $\mathcal{S}$ is a set of states. 
This can be formalized as a Markov decision process (MDP) which is defined as the tuple $M = \left( \mathcal{S}, \mathcal{A}, \mathcal{P}, \mathcal{R}, \gamma  \right)$. Here $\mathcal{P}(s_{t+1} \vert s_t, a_t)$ is the transition probability to reach the next state  $s_{t+1}$ given the current state $s_t$ and action $a_t$, and $\gamma \in [0,1)$ is a discount factor.
RL encompasses a variety of approaches~\cite{sutton2018reinforcement}, which can be categorized into model-free approaches like proximal policy optimization~\cite{schulman2017proximal} and model-based techniques such as model predictive control~\cite{GARCIA1989335}. 

In this work, we use the RL algorithm MuZero~\cite{Schrittwieser_2020} for circuit generation. The general procedure, however, can be implemented using other RL algorithms, too. MuZero uses a representation function, which is approximated by a neural network to transform the observation $o_t \in \mathcal{O}$ into a latent state representation $s_t$, which leads to a partially observable MDP $M = \left( \mathcal{S}, \mathcal{A}, \mathcal{P}, \mathcal{R}, \mathcal{O}, \gamma  \right)$, where $\mathcal{O}$ is the observation space. 

To apply an RL algorithm to circuit generation, the RL agent places gates sequentially, where the set of actions $\mathcal{A}$ corresponds to the predefined set of unitary gates $\mathcal{V}$ [cf. discussion below Eq.~\eqref{eq:encoding_circuits}]. The reward $r_t$ is determined by the performance of a QML model that utilizes the data encoding circuit $U_t$ at time $t$, i.e., the observations corresponds to the configurations of the quantum encoding circuit $o_t = U_t$.

At each time step $t$, the model performs a Monte Carlo tree search and processes the hidden state for $k = 1,\dots,K$ steps. The model then uses neural networks to approximate an internal reward, an internal value function, and internal policies.
At the end of each Monte Carlo tree search step, an action is sampled from the policy $\Pi_t(a_{t+k+1} \vert o_1,\dots,o_t,a_{t+1},\dots,a_{t+k})$, and the actual reward and value function are computed. The weights of the three networks are optimized during the training process to minimize the difference between the internal hidden state predictions and real outcomes. For more detail on the algorithm, we refer to~\cite{Schrittwieser_2020}.

Using MuZero for circuit generation offers several advantages over other potential RL algorithms. Treating the problem as a partially observable MDP and transforming the observation space into the latent space of a neural network reduces the amount of knowledge needed about the underlying dynamics. This makes the algorithm particularly well suited for QML, where a comprehensive understanding of the complicated dynamics between the encoding circuit and the data remains incomplete. Another critical advantage is the algorithm's scalability in computational resources. The algorithm approximates all necessary functions using classical neural networks, which means that the number of actual circuit evaluations, which typically are the primary computational bottleneck of a QML model, are reduced.
In the next section, we provide a detailed explanation on how MuZero can be utilized to generate encoding circuits that achieve good performance.

\subsection{Encoding circuit generation} \label{sec:circuit_gen}

\begin{figure*}[t]
	\centering
	\includegraphics[scale=0.8]{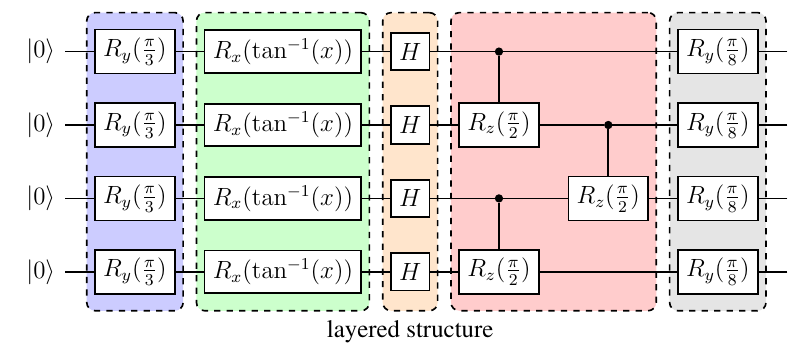}
	\caption{Example of a circuit with a layered structure with $q=4$ qubits that has been created after 5 actions of the RL agent. The operators are chosen from the action set described in Tab.~\ref{tab: action_set}.
 }
	\label{fig: layered_circ}
\end{figure*}

The workflow of the circuit generation process based on MuZero, denoted in the following as MCS, is shown in Fig.~\ref{fig: concept_layout}. The RL agent constructs quantum circuits using a layered structure where each gate operator is applied to all accessible qubits. Two-qubit gates are applied in a linear nearest-neighbor configuration. Figure~\ref{fig: layered_circ} shows a typical example of such a layered circuit. Although a more flexible structure is generally possible, full flexibility poses a major challenge due to the exponentially growing search space, which makes it difficult to find optimal solutions. This is confirmed by our benchmarks in Sec.~\ref{sec: numerical results} which show a better performance of the layered structure compared to its more flexible counterpart.

The action set of the RL algorithm chosen for the remainder of the manuscript is listed in Tab.~\ref{tab: action_set}. While additional one and multi-qubit gates could be considered, we limit the selection to those primarily used in encoding circuits for QML to reduce the search space. Rotational gates are employed to encode input features or manipulate the quantum state. Besides linear data encoding, we also include gates with nonlinear transformations of the data using $\arctan(x)$.
In our approach, we use fixed circuit parameters instead of trainable ones. This allows the MCS to optimize the parameters implicitly by selecting gates that lead to an optimal reward. The fixed parameters are given by $\pi/n$ with $n \in \{1,2,3,4,8\}$. 
The total number of qubits must be predefined and, in our implementation, is always the number of features of the ML problem, ensuring that each feature is represented once in every layer. While it is conceptually reasonable to allow the agent to choose the number of qubits, we do not consider this to reduce the complexity of the actions. In general, the action set can be adapted to match the hardware-specific native gate set.

The environment of the RL agent is initialized by an empty circuit with all qubits in the zero state. The agent then iteratively adds gates until a specified number $n_\text{maxdepth}$ of layers has been reached, which triggers a re-initialization to an empty circuit. 
Furthermore, we address the exploration-exploitation dilemma by incorporating random initial architectures to improve the agent's exploration capabilities. These random restarts are executed with a probability of $20\%$, and they are achieved by randomly picking $ \nicefrac{n_\text{maxdepth}}{2}$ actions and ensuring that at least one feature encoding layer is included. 

The reward the agent receives is given by
\begin{equation}
     r_t(c_t) =
    \begin{cases}
    r_{g} &\text{ \, if \,} c_t > c_g^* \\
    r_{l} &\text{ \, if \,} \mathrm{c}_* \geq c_t > c_l^* \\
   r_{p_1} &\text{ \, if \,} c_t \leq c_l^*\\
   r_{p_2} &\text{ \, if \,}  c_{t} = -1\\
    \end{cases}\,,
\end{equation}
Here, $c_t$ denotes the cross-validation score at time $t$. The specific metric that is used to calculate $c_t$ may vary depending on the type of problem. The agent is granted a reward $r_g$ for achieving a global success, i.e., surpassing the highest score $\mathrm{c}_g^*$ stored over all time steps. Additionally, a reward $r_l$ is given for local improvements, i.e., when the highest score $c_l^*$ of the current circuit is exceeded.
A negative reward $r_{p_1}$ is applied if the score does not change by adding an additional operator, discouraging redundancies and encouraging the development of shorter circuits.
Another negative reward $r_{p_2}$ is imposed if the encoding circuit $U_t$ does not contain any gates dependent on the data $\xdata$. In such cases, the cross-validation evaluation is skipped and we manually set $c_t = -1$. In our implementations, we typically use $\abs{r_g} > \abs{r_l} > \abs{r_{p_1}} \approx \abs{r_{p_2}}$.

In principle, the presented MCS algorithm can be used with a variety of QML models. For practical reasons, we focus exclusively on QML models based on PQKs as they provide a robust QML framework that is quick to evaluate. 
PQKs use quantum states as an intermediate step and project back to classical representations using observables before evaluating the kernel. This process can be viewed as pre-processing the input data with the quantum model before applying a classical kernel method. We employ a Gaussian outer kernel with 1-reduced-density matrices as proposed in~\cite{Huang_2021}
\begin{equation}
\begin{aligned}
K^{\rm PQK}(\xdata,\xdata') =  & \\
 & \hspace{-6.7em} \exp\Bigg(-\gamma \sum_{k} \sum_{O \in \{X_k, Y_k, Z_k\}} \big[f_{O,k}(\xdata) - f_{O,k}(\xdata')\big]^2 \Bigg)\,.
\end{aligned}
\label{eq:gaussian_pqk}
\end{equation}
Here, $\gamma > 0$ is a hyperparameter and $f_{O,k} = \tr(\rho(\xdata) O_k)$, where the observable $O_k$ is taken from the set of single-qubit Pauli matrices $\{X, Y, Z\}$ and applied to qubit $k$. The resulting kernel is then used in a support vector machine (SVM) to create a QSVM.

We primarily chose PQKs for two reasons: they are efficient to evaluate and have shown good results for various ML tasks. Unlike fidelity quantum kernels that are based on the overlap of the two quantum states~\cite{Havl_ek_2019}, the evaluation of quantum circuits with PQKs scales linearly with the data set size. Additionally, caching of intermediate results enables a fast evaluation of the cross-validation score. 
We did not choose QNNs here because, in contrast to PQKs, QNNs are parameterized and require an expensive training routine that involves numerous evaluations of quantum circuits for each training data point, making the search for optimal architectures prohibitive. The implementation of the circuits, the PQK, and the QML models is achieved using the library \emph{sQUlearn}~\cite{kreplin2023squlearn}.

\begin{table}[t]
\caption{Action set, i.e., set of gates which can be chosen as actions by the RL agent. Gates are added as layers to all qubits at the end of the circuit. Two-qubit gates are applied in a linear nearest neighbor entangling configuration. Gates containing an $n$ in the argument are available in $5$ variants with $n \in \{1,2,3,4,8\}$.}
\begin{tabularx}{\linewidth}{@{}XXXXX@{}}
\toprule
$X$              & $CX$  & $R_x(\pi x)$  & $R_x(\frac{\pi}{n})$ \\[1mm]
$Y$ & $CY$ & $R_y(\pi x)$ & $R_y(\frac{\pi}{n})$ \\[1mm]
$Z$ & $CZ$ & $R_z(\pi x)$ & $R_z(\frac{\pi}{n})$ \\[1mm]
$H$ & $CR_x(\frac{\pi}{n})$ & $CR_y(\frac{\pi}{n})$ & $CR_z(\frac{\pi}{n})$ \\[1mm]
$R_x( \arctan(x))$ & $R_y( \arctan(x))$ & $R_z( \arctan(x))$ & \\[0mm] \bottomrule
\end{tabularx}

\label{tab: action_set}
\end{table}

\section{Results}
\label{sec: results}

To assess the effectiveness of the MCS algorithm, we evaluate the performance of QML models based on its generated circuits against models using reference encoding circuits from the literature~\cite{Hubreg_, peters2021machine, Haug_2023, Havl_ek_2019}. Additionally, we contrast these outcomes with circuits generated using a genetic algorithm~\cite{Altares_Lopez_2021} and circuits constructed from randomly chosen gate sequences. Furthermore, we compare these results to various classical machine learning models. Our evaluation is conducted across three distinct datasets, encompassing two regression tasks and one classification task.

\subsection{Benchmark set-up\label{sec: benchmark}}

This section details the set-up of our benchmark calculations, which are consistent across all compared methods and circuits.
First, the data set is split into two parts: a training set and a test set. The training set is used in cross-validation to determine suitable encoding circuits and to optimize the hyperparameters of the underlying (Q)ML model. The final score with which the different approaches are compared is calculated with the test set exclusively.
The architecture search algorithms, including the MCS algorithm, random searches, and the genetic algorithm, follow a three-step procedure:
\begin{enumerate}
    \item Generation and selection: We generate various circuits and select the five best circuits based on the cross-validation score on the training data.
    \item Hyperparameter optimization: We optimize the hyperparameters of the underlying (Q)ML model using the cross-validation score of the training data.
    \item Final evaluation: We train the final model on the full training data and evaluate the final score using the test data.
\end{enumerate}
The cross-validation in steps (1) and (2) uses the mean squared error (MSE) for regression tasks and the accuracy for the classification example, based on five fixed folds for every data set. 
The hyperparameters in step (1) are consistent across all architecture search algorithms. 
The hyperparameter optimization in step (2) is performed using the Optuna library~\cite{optuna_2019}, optimizing both the hyperparameters of the (Q)ML model and the hyperparameter $\gamma$ of the PQK (if applicable).
The number of executed cross-validations is comparable for all architecture search algorithms and is in the order of $5,000$ evaluations.
For the reference circuits from the literature, step (1) is replaced by optimizing the variational parameters of the PQCs with target alignment (cf. Sec. \ref{sec: reference_circs}).
For classical ML models, step (1) is not required.
All results in this work are obtained using noiseless statevector simulations.
In the following, we discuss details specific to each method.

\subsubsection{MCS}
We limit the maximum circuit depth for the MCS algorithm to $n_\text{maxdepth} = 10$ layers to maintain hardware efficiency.
The reward values are set to $r_g = 100$, $r_l = 50$, and $r_{p_1} = r_{p_2} = -1$.
The final encoding circuits are obtained from those generated during both the training and the inference phases of the MuZero search algorithm. In the latter, the weights of the neural networks in the MuZero algorithm remain fixed.

\subsubsection{Random encoding circuits}
\label{sec: random_circs}

Random circuits are generated from the same gate set as used by the MCS (cf. Tab. ~\ref{tab: action_set}). Using the same layered approach as shown in Fig.~\ref{fig: layered_circ} and detailed in Section~\ref{sec:circuit_gen}, gate-layers are randomly selected with a random circuit depth between 2 and $n_\text{maxdepth}=10$.

Additionally, we generate circuits that do not adhere to a layer-wise application of gates. In this fully random approach, one- and two-qubit gates are placed at every possible position within the circuit. The rotation angles of individual gates are randomly sampled from a uniform distribution between $[0,2\pi]$. The maximum number of placed gates matches the total number of gates in the layered approach with the maximum depth.

In both approaches, we ensure that all features in $\xdata$ are encoded at least once in the circuits. This slightly favors the randomly generated circuits compared to the circuits generated by the MCS because including data-dependent gates in the circuit is something that the MCS algorithm has to learn during training. For both random generation methods, we sample $5,000$ circuits and select the top $5$ circuits based on the cross-validation score of the training data.

\subsubsection{Genetic algorithms}
\label{sec: genetic_circs}

We generate circuits using a slightly modified version of the genetic algorithm proposed in~\cite{Altares_Lopez_2021}. While~\cite{Altares_Lopez_2021} employs fidelity-based quantum kernels, we integrate PQKs into the algorithm. Furthermore, we adjust the training procedure to use the cross-validation score, departing from the original methodology that relies on training and testing scores, in order to align with the RL and random search methods.
As in its original proposal, the genetic algorithm is not bound to the layered circuit structure. It has a high level of flexibility when building circuits, e.g., it can encode multiple features on one qubit wire, as well as having full freedom in placing entangling gates.

\subsubsection{Reference circuits}
\label{sec: reference_circs}

We benchmark our generated circuits against five effective circuits from the literature~\cite{Hubreg_, peters2021machine, Haug_2023, Havl_ek_2019, kreplin2023reduction}.
Our selection includes a data-reuploading scheme that has been successful in previous QML studies~\cite{Hubreg_, bowles2024better}.  With the exception of one circuit, all circuits feature variational parameters $\pmb{\theta}$ that we optimize by maximizing the kernel-target-alignment metric over the training set~\cite{Hubreg_, Glick2021hdc}. The chosen reference circuits provide a selection of common strategies found in the literature, encompassing design principles such as hardware-efficient entanglement and the placement of trainable variational parameters within the circuits and the encoding. The exact circuits can be found in Appendix~\ref{app:ref_circuits}.

\subsubsection{Classical models}
\label{sec: cml}
As classical models, we use support vector machines, neural networks, random forests, gradient boosting, and kernel ridge regression for regression tasks.
Gradient boosting is implemented using the XGBoost library~\cite{Chen_2016}, while neural networks are implemented using TensorFlow~\cite{tensorflow2015whitepaper}. The remaining models are implemented using scikit-learn~\cite{scikitlearn}. All classical kernel methods are based on radial basis function kernels. 

\subsubsection{Data sets}
\label{sec: data set}
The data sets have been chosen for their QML compatibility, which imposes restriction on the size and dimensionality of the data. Additionally, we aim for selecting difficult data sets (or data set variants).

The \emph{Quantum fashion MNIST} (QFMNIST) data set is based on the fashion-MNIST data set~\cite{xiao2017fashionmnist}, and prepared in a quantum state as described in~\cite{Huang_2021, Jerbi_2023}.
The features are encoded into quantum Hilbert space as proposed by Havlicek et
al.~\cite{Havl_ek_2019}. We ommit the subsequent arbitrary single qubit rotations applied in~\cite{Jerbi_2023}, which simplifies the problem for classical methods. Nevertheless, we maintain a high level of difficulty by selecting only the two most challenging classes from the fashion-MNIST dataset.
This results in a regression problem using an artificial quantum data set. The data points are standardized and the original features are reduced using scikit-learn's PCA implementation~\cite{scikitlearn} using the first $8$ principal components as features. 
We use $n_{\text{training}} = 1000$ training points, and $n_{\text{test}} = 200$ test points with target values scaled in a range from $[-1,1]$.

The \emph{California housing} data set from scikit-learn~\cite{scikitlearn} constitutes a regression problem that aims to predict the prices of houses in California. The data is normalized, and we choose a fixed sub set of $n_{\text{data}} = 1,000$ randomly chosen data points. These data points are partitioned into $n_{\text{training}} = 700$ training points and $n_{\text{test}} = 300$ test points. The data set has $8$ features.

The \emph{Two-curves (diff)} data set~\cite{bowles2024better} is a classification task which models the curvature and the minimum distance of two curves in a $d$-dimensional space. Here, we follow the data creation procedure described in~\cite{bowles2024better}.
The problem's difficulty can vary depending on the choice of degree $d$ and the number of features. We use the most difficult version of the data set described in~\cite{bowles2024better} using $10$ features and a degree of $d=20$. The data is standardized, and we use $n_{\text{data}} = 300$ samples and split them into $n_{\text{training}} = 240$ training points for cross-validation, and $n_{\text{test}} = 60$ test points. 

\subsection{Experiments}
\label{sec: numerical results}
\begin{figure*}[t]
    \centering
    \includegraphics[width=0.49\linewidth]{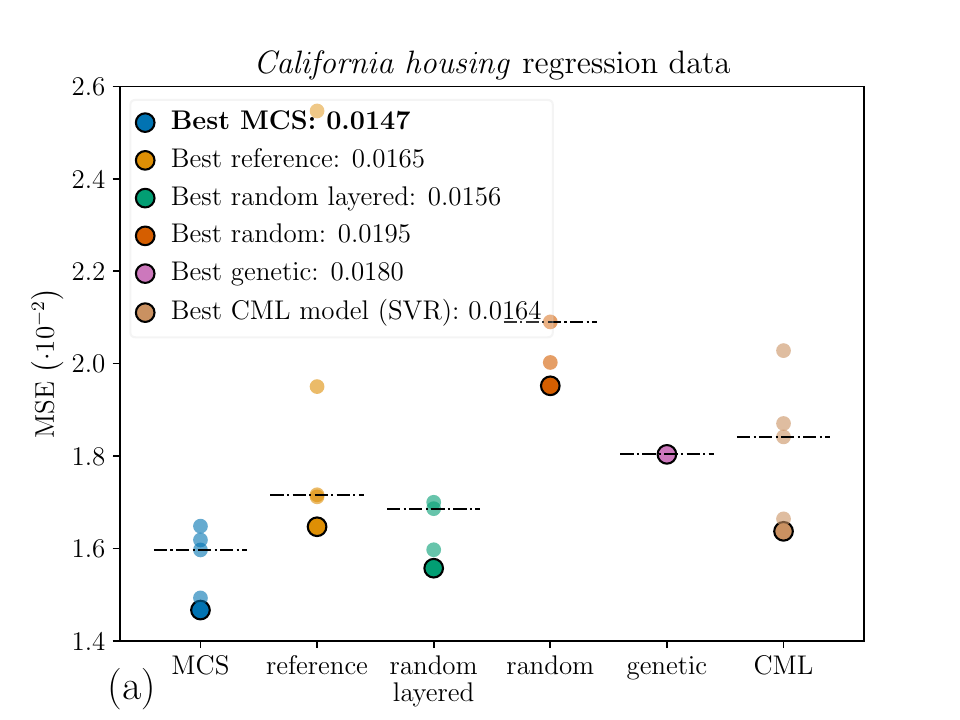}
    \hfill
    \includegraphics[width=0.49\linewidth]{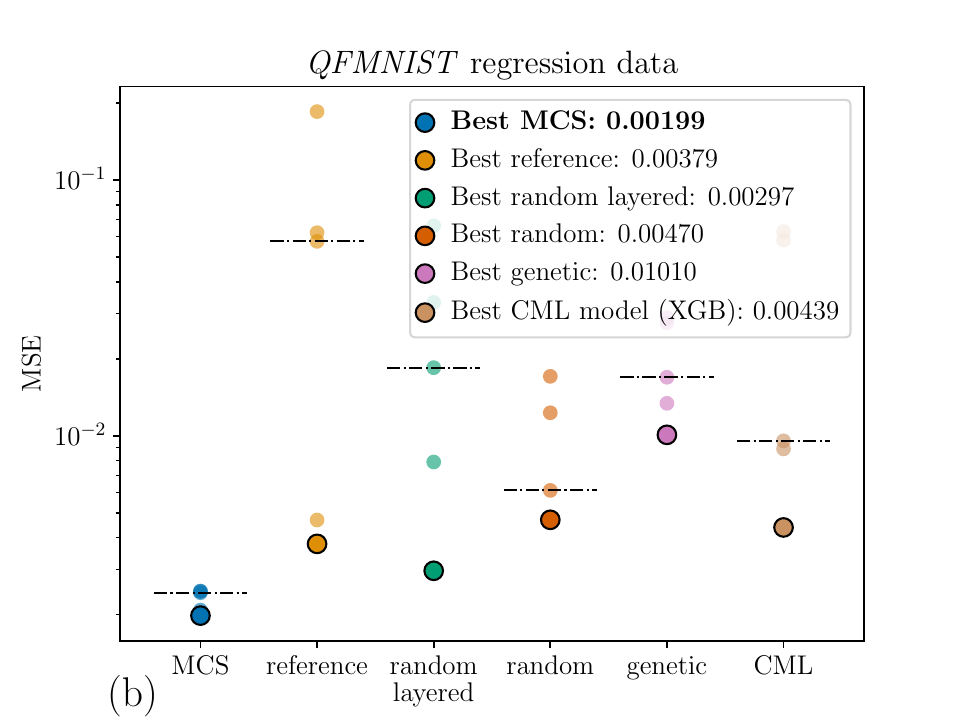}
    
    \caption{Performance on the regression tasks. Results are shown for the California housing data (a) and the QFMNIST data (b). The first category marks the results of a quantum support vector regression (QSVR) based on encoding circuits built by the MCS (Sec.~\ref{sec:circuit_gen}). The second category shows the results of a QSVR based on a variety of reference encoding circuits (Sec.~\ref{sec: reference_circs}). The third and fourth columns mark the results of a QSVR based on random encoding circuits (Sec.~\ref{sec: random_circs}).
    The fifth category shows the results of an encoding circuit built by a genetic algorithm (Sec.~\ref{sec: genetic_circs}).
    The category CML shows the results of different classical ML models, as described in Sec.~\ref{sec: cml}. The dashed lines mark the median performance of each category.}
    \label{fig: combined_results}
\end{figure*}

Figure~\ref{fig: combined_results} shows the MSE of the two regression data sets. For the California housing data set Fig.~\ref{fig: combined_results}(a) the best performance is achieved by the models based on the encoding circuits generated by the MCS algorithm. The best classical model is the SVR, with an $\mathrm{MSE}_{\text{SVR}} = 0.0164$ test score. Overall, the classical models have the largest performance variance, which is not surprising given that the set of classical models used here is very diverse. 

We observe a similar behavior for the \mbox{QFMNIST} regression problem.
As shown in Fig.~\ref{fig: combined_results}(b), the best performance is achieved with the circuits found by the MCS algorithm. The five different circuits produced by the MCS perform similarly well. As a result, the differences cannot be distinguished in the plot at this scale. This demonstrates the MCS algorithm's ability to create robust circuit architectures that consistently yield good results. The models using reference circuits yield solid solutions, too. However, some of the chosen reference circuits lead to significantly underperforming models. 
In contrast to Fig.~\ref{fig: combined_results}(a), in which the best reference circuit from the literature has no trainable parameters \cite{peters2021machine}, 
the same circuit in Fig.~\ref{fig: combined_results}(b) shows the worst results among all encoding circuits considered. Having the same circuit perform the best on one problem and the worst on another highlights the need for a tailored approach. The better performance of most QML models compared to the best classical ML model on the QFMNIST data is likely attributed to the dataset's synthetic nature, specifically designed to be difficult to simulate classically~\cite{Huang_2021}.

\begin{figure}[t]
    \includegraphics[width=1.05\linewidth]{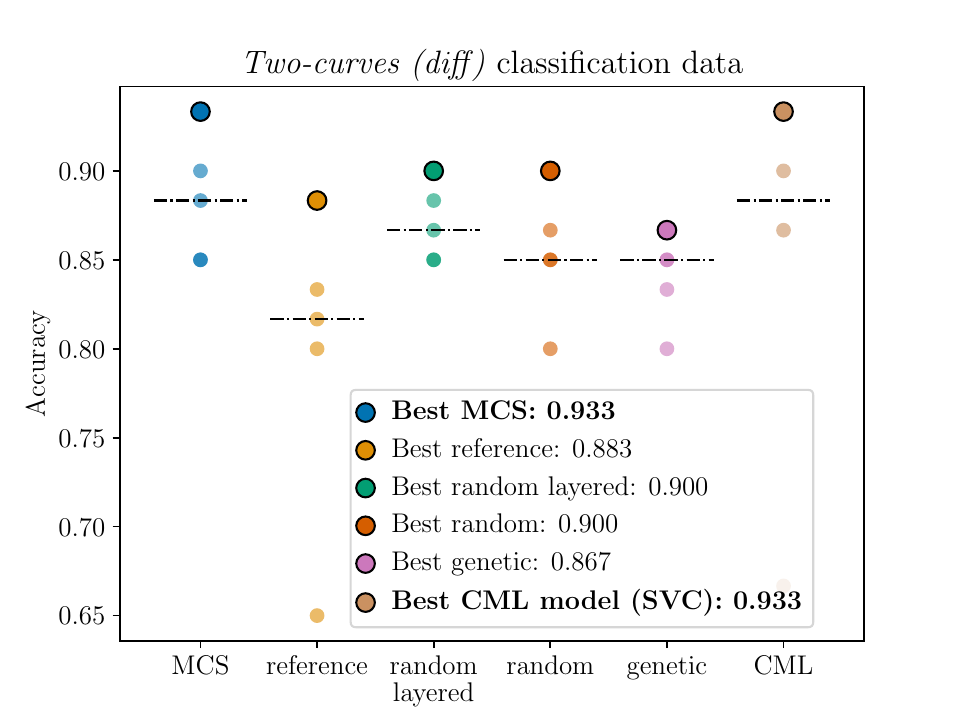}
    \caption{Performance on the two-curves (diff) classification task. The columns follow the description in Fig.~\ref{fig: combined_results}, with the QSVR models replaced by quantum support vector classification models, and dotted lines as median performance.}
    \label{fig: twocurves_results}
\end{figure}

The accuracy achieved by the models on the classification task on the two-curves (diff) data set is shown in Fig.~\ref{fig: twocurves_results}.
For this problem, the QSVM model with encoding circuits generated with the MCS algorithm, and the classical SVM perform equally well with a test accuracy of $0.933$.
The reference circuits perform slightly worse than the random circuits, with the best reference architecture for this problem being the optimized approach from~\cite{Haug_2023} (\ref{fig: ref circuits}(c)).

Comparing the results across all data sets, we found that the layered structure is generally more favorable compared to full flexibility. This finding holds true when comparing both the median performance and the best-case performance.
Not only does this apply when comparing the genetic algorithm, which lags behind in all cases, to the MCS, but also when comparing the two random search approaches, both with and without the layered structure. This serves as confirmation for our choice of the layered circuit structure.

An important insight from all experiments comes from comparing models based on reference circuits with those using circuits found by random search approaches. While the competitive performance of models using circuits generated by MCS is expected due to their problem-tailoring, it is noteworthy that random searches can also yield high-performance circuits. The QSVMs using random encoding circuits consistently outperform the reference encoding circuits. Since the randomly generated circuits have been post-selected on their performance on a cross-validation score it is maybe not surprising that this approach can yield satisfactory circuits. It is noteworthy, however, that the number of well performing circuits generated in this fashion were quite high, which indicates that there are often multiple equivalent circuits that lead to similar results. Overall this confirms that by moving beyond heurstics towards tailored circuits, the performance of QML models can be increased.

\section{Discussion \label{sec:discussion}}

In this study, we demonstrated the automatic generation of encoding circuits for QML using the MCS algorithm. Our experiments illustrated the effectiveness of this method in creating tailored circuits that considerably improve the performance of the QML model. The best quantum model in all of our benchmarks was always based on a circuit constructed by the MCS algorithm. This approach benefits from learning from the problem structure, enabling the creation of tailored circuits that outperform other methods. This underscores the main claim of this work: contrary to common practice in QML research, where heuristic circuit architectures are used and their parameters optimized, better results can be achieved with custom circuits. 

Interestingly, random search strategies also produce good results, often matching or even surpassing the performance of reference circuits from the literature. This is particularly notable given that in this work, the trainable parameters of the reference circuits have been optimized, while the parameters in the random search have been chosen randomly. This raises the question of how much parameter optimization in encoding circuits really contributes to improving the performance of QML models based on PQKs.

Although our experiments show better performance of QML models in two out of three cases, we do not claim that they are superior to classical models. Although the benchmarking problems in our study were selected with care, they are limited when it comes to the number of data points and the dimensionality. Predicting scaling behavior for larger problems is an open question in this context. 

Looking to the future, there are several opportunities to extend the work presented in this study. 
The MCS algorithm can be adapted to generate circuits for different QML models, such as QNNs, and could serve as a general circuit optimization tool for other circuit approaches outside of QML.
Furthermore, incorporating operators that reflect the symmetries of the data set could introduce an informed inductive bias, leading to more practical circuit designs.
Moreover, the MCS is promising for transferring the learned RL model from one data set to another, since the hidden state representation of the circuit has already been learned. Further research could evaluate the ability to generalize to new data sets without requiring intensive retraining of the RL algorithm.

In conclusion, our study highlights the significant potential of using RL for the generation of problem-specific encoding circuits for QML. The results show that tailored circuits can significantly improve the performance of QML models and provide a solid foundation for future research and development in the field of QML.

\backmatter
\bmhead{Acknowledgments}
We thank Jan Schnabel, Tobias Nagel and Philipp Wagner for insightful discussions. This work was supported by the Baden-Württemberg Ministry of Economic Affairs, Labor, and Tourism in the frame of the Competence Center Quantum Computing Baden-Württemberg (project SEQUOIA End-to-End).

\bibliography{muzero_qml.bib}

\begin{thebibliography}{10}
\expandafter\ifx\csname url\endcsname\relax
  \def\url#1{\burl{#1}}\fi
\expandafter\ifx\csname urlprefix\endcsname\relax\def\urlprefix{}\fi
\providecommand{\bibinfo}[2]{#2}
\providecommand{\doi}[1]{\url{https://doi.org/#1}}
\bibcommenthead

\bibitem{Preskill_2018}
\bibinfo{author}{Preskill, J.}
\newblock \bibinfo{title}{Quantum computing in the {NISQ} era and beyond}.
\newblock \emph{\bibinfo{journal}{Quantum}} \textbf{\bibinfo{volume}{2}},
  \bibinfo{pages}{79} (\bibinfo{year}{2018}).

\bibitem{cerezo2022qml}
\bibinfo{author}{Cerezo, M.}, \bibinfo{author}{Verdon, G.},
  \bibinfo{author}{Huang, H.-Y.}, \bibinfo{author}{Cincio, L.} \&
  \bibinfo{author}{Coles, P.}
\newblock \bibinfo{title}{Challenges and opportunities in quantum machine
  learning}.
\newblock \emph{\bibinfo{journal}{Nature Computational Science}}
  \textbf{\bibinfo{volume}{2}} (\bibinfo{year}{2022}).

\bibitem{mitarai2018}
\bibinfo{author}{Mitarai, K.}, \bibinfo{author}{Negoro, M.},
  \bibinfo{author}{Kitagawa, M.} \& \bibinfo{author}{Fujii, K.}
\newblock \bibinfo{title}{Quantum circuit learning}.
\newblock \emph{\bibinfo{journal}{Phys. Rev. A}} \textbf{\bibinfo{volume}{98}},
  \bibinfo{pages}{032309} (\bibinfo{year}{2018}).

\bibitem{Benedetti_2019}
\bibinfo{author}{Benedetti, M.}, \bibinfo{author}{Lloyd, E.},
  \bibinfo{author}{Sack, S.} \& \bibinfo{author}{Fiorentini, M.}
\newblock \bibinfo{title}{Parameterized quantum circuits as machine learning
  models}.
\newblock \emph{\bibinfo{journal}{Quantum Science and Technology}}
  \textbf{\bibinfo{volume}{4}}, \bibinfo{pages}{043001} (\bibinfo{year}{2019}).

\bibitem{Schuld_2019}
\bibinfo{author}{Schuld, M.} \& \bibinfo{author}{Killoran, N.}
\newblock \bibinfo{title}{Quantum machine learning in feature hilbert spaces}.
\newblock \emph{\bibinfo{journal}{Physical Review Letters}}
  \textbf{\bibinfo{volume}{122}} (\bibinfo{year}{2019}).

\bibitem{Havl_ek_2019}
\bibinfo{author}{Havl{\'{\i}}{\v{c}}ek, V.} \emph{et~al.}
\newblock \bibinfo{title}{Supervised learning with quantum-enhanced feature
  spaces}.
\newblock \emph{\bibinfo{journal}{Nature}} \textbf{\bibinfo{volume}{567}},
  \bibinfo{pages}{209--212} (\bibinfo{year}{2019}).

\bibitem{Huang_2021}
\bibinfo{author}{Huang, H.-Y.} \emph{et~al.}
\newblock \bibinfo{title}{Power of data in quantum machine learning}.
\newblock \emph{\bibinfo{journal}{Nature Communications}}
  \textbf{\bibinfo{volume}{12}} (\bibinfo{year}{2021}).

\bibitem{TILLY20221}
\bibinfo{author}{Tilly, J.} \emph{et~al.}
\newblock \bibinfo{title}{The variational quantum eigensolver: A review of
  methods and best practices}.
\newblock \emph{\bibinfo{journal}{Physics Reports}}
  \textbf{\bibinfo{volume}{986}}, \bibinfo{pages}{1--128}
  (\bibinfo{year}{2022}).
\newblock \bibinfo{note}{The Variational Quantum Eigensolver: a review of
  methods and best practices}.

\bibitem{Peruzzo2014}
\bibinfo{author}{Peruzzo, A.} \emph{et~al.}
\newblock \bibinfo{title}{A variational eigenvalue solver on a photonic quantum
  processor}.
\newblock \emph{\bibinfo{journal}{Nature Communications}}
  \textbf{\bibinfo{volume}{5}}, \bibinfo{pages}{4213} (\bibinfo{year}{2014}).

\bibitem{farhi2014quantum}
\bibinfo{author}{Farhi, E.}, \bibinfo{author}{Goldstone, J.} \&
  \bibinfo{author}{Gutmann, S.}
\newblock \bibinfo{title}{A quantum approximate optimization algorithm}
  (\bibinfo{year}{2014}).

\bibitem{ragone2023unified}
\bibinfo{author}{Ragone, M.} \emph{et~al.}
\newblock \bibinfo{title}{A unified theory of barren plateaus for deep
  parametrized quantum circuits} (\bibinfo{year}{2023}).

\bibitem{larocca2024review}
\bibinfo{author}{Larocca, M.} \emph{et~al.}
\newblock \bibinfo{title}{A review of barren plateaus in variational quantum
  computing} (\bibinfo{year}{2024}).

\bibitem{kuebler2021inductive}
\bibinfo{author}{K\"{u}bler, J.}, \bibinfo{author}{Buchholz, S.} \&
  \bibinfo{author}{Sch\"{o}lkopf, B.}
\newblock \bibinfo{editor}{Ranzato, M.}, \bibinfo{editor}{Beygelzimer, A.},
  \bibinfo{editor}{Dauphin, Y.}, \bibinfo{editor}{Liang, P.} \&
  \bibinfo{editor}{Vaughan, J.~W.} (eds) \emph{\bibinfo{title}{The inductive
  bias of quantum kernels}}.
\newblock (eds \bibinfo{editor}{Ranzato, M.}, \bibinfo{editor}{Beygelzimer,
  A.}, \bibinfo{editor}{Dauphin, Y.}, \bibinfo{editor}{Liang, P.} \&
  \bibinfo{editor}{Vaughan, J.~W.}) \emph{\bibinfo{booktitle}{Advances in
  Neural Information Processing Systems}}, Vol.~\bibinfo{volume}{34},
  \bibinfo{pages}{12661--12673} (\bibinfo{publisher}{Curran Associates, Inc.},
  \bibinfo{year}{2021}).

\bibitem{585893}
\bibinfo{author}{Wolpert, D.} \& \bibinfo{author}{Macready, W.}
\newblock \bibinfo{title}{No free lunch theorems for optimization}.
\newblock \emph{\bibinfo{journal}{IEEE Transactions on Evolutionary
  Computation}} \textbf{\bibinfo{volume}{1}}, \bibinfo{pages}{67--82}
  (\bibinfo{year}{1997}).

\bibitem{bowles2023contextuality}
\bibinfo{author}{Bowles, J.}, \bibinfo{author}{Wright, V.~J.},
  \bibinfo{author}{Farkas, M.}, \bibinfo{author}{Killoran, N.} \&
  \bibinfo{author}{Schuld, M.}
\newblock \bibinfo{title}{Contextuality and inductive bias in quantum machine
  learning} (\bibinfo{year}{2023}).

\bibitem{ragone2023representation}
\bibinfo{author}{Ragone, M.} \emph{et~al.}
\newblock \bibinfo{title}{Representation theory for geometric quantum machine
  learning} (\bibinfo{year}{2023}).

\bibitem{PRXQuantum3030341}
\bibinfo{author}{Larocca, M.} \emph{et~al.}
\newblock \bibinfo{title}{Group-invariant quantum machine learning}.
\newblock \emph{\bibinfo{journal}{PRX Quantum}} \textbf{\bibinfo{volume}{3}},
  \bibinfo{pages}{030341} (\bibinfo{year}{2022}).

\bibitem{PRXQuantum4010328}
\bibinfo{author}{Meyer, J.~J.} \emph{et~al.}
\newblock \bibinfo{title}{Exploiting symmetry in variational quantum machine
  learning}.
\newblock \emph{\bibinfo{journal}{PRX Quantum}} \textbf{\bibinfo{volume}{4}},
  \bibinfo{pages}{010328} (\bibinfo{year}{2023}).

\bibitem{Schatzki2022tfq}
\bibinfo{author}{Schatzki, L.}, \bibinfo{author}{Larocca, M.},
  \bibinfo{author}{Nguyen, Q.~T.}, \bibinfo{author}{Sauvage, F.} \&
  \bibinfo{author}{Cerezo, M.}
\newblock \bibinfo{title}{{Theoretical guarantees for permutation-equivariant
  quantum neural networks}}.
\newblock \emph{\bibinfo{journal}{npj Quantum Inf.}}
  \textbf{\bibinfo{volume}{10}}, \bibinfo{pages}{12} (\bibinfo{year}{2024}).

\bibitem{gili2023inductive}
\bibinfo{author}{Gili, K.}, \bibinfo{author}{Alonso, G.} \&
  \bibinfo{author}{Schuld, M.}
\newblock \bibinfo{title}{An inductive bias from quantum mechanics: learning
  order effects with non-commuting measurements} (\bibinfo{year}{2023}).

\bibitem{Schrittwieser_2020}
\bibinfo{author}{Schrittwieser, J.} \emph{et~al.}
\newblock \bibinfo{title}{Mastering atari, go, chess and shogi by planning with
  a learned model}.
\newblock \emph{\bibinfo{journal}{Nature}} \textbf{\bibinfo{volume}{588}},
  \bibinfo{pages}{604–609} (\bibinfo{year}{2020}).

\bibitem{Swiechowski2023}
\bibinfo{author}{Świechowski, M.}, \bibinfo{author}{Godlewski, K.},
  \bibinfo{author}{Sawicki, B.} \& \bibinfo{author}{Mańdziuk, J.}
\newblock \bibinfo{title}{Monte carlo tree search: a review of recent
  modifications and applications}.
\newblock \emph{\bibinfo{journal}{Artificial Intelligence Review}}
  \textbf{\bibinfo{volume}{56}}, \bibinfo{pages}{2497--2562}
  (\bibinfo{year}{2023}).

\bibitem{Rebentrost_2014}
\bibinfo{author}{Rebentrost, P.}, \bibinfo{author}{Mohseni, M.} \&
  \bibinfo{author}{Lloyd, S.}
\newblock \bibinfo{title}{Quantum support vector machine for big data
  classification}.
\newblock \emph{\bibinfo{journal}{Physical Review Letters}}
  \textbf{\bibinfo{volume}{113}} (\bibinfo{year}{2014}).

\bibitem{Altares_Lopez_2021}
\bibinfo{author}{Altares-López, S.}, \bibinfo{author}{Ribeiro, A.} \&
  \bibinfo{author}{García-Ripoll, J.~J.}
\newblock \bibinfo{title}{Automatic design of quantum feature maps}.
\newblock \emph{\bibinfo{journal}{Quantum Science and Technology}}
  \textbf{\bibinfo{volume}{6}}, \bibinfo{pages}{045015} (\bibinfo{year}{2021}).

\bibitem{Pozzi.2022}
\bibinfo{author}{Pozzi, M.~G.}, \bibinfo{author}{Herbert, S.~J.},
  \bibinfo{author}{Sengupta, A.} \& \bibinfo{author}{Mullins, R.~D.}
\newblock \bibinfo{title}{Using reinforcement learning to perform qubit routing
  in quantum compilers}.
\newblock \emph{\bibinfo{journal}{ACM Transactions on Quantum Computing}}
  \textbf{\bibinfo{volume}{3}} (\bibinfo{year}{2022}).

\bibitem{Zhang_2022}
\bibinfo{author}{Zhang, S.-X.}, \bibinfo{author}{Hsieh, C.-Y.},
  \bibinfo{author}{Zhang, S.} \& \bibinfo{author}{Yao, H.}
\newblock \bibinfo{title}{Differentiable quantum architecture search}.
\newblock \emph{\bibinfo{journal}{Quantum Science and Technology}}
  \textbf{\bibinfo{volume}{7}}, \bibinfo{pages}{045023} (\bibinfo{year}{2022}).

\bibitem{Zhang_2021}
\bibinfo{author}{Zhang, S.-X.}, \bibinfo{author}{Hsieh, C.-Y.},
  \bibinfo{author}{Zhang, S.} \& \bibinfo{author}{Yao, H.}
\newblock \bibinfo{title}{Neural predictor based quantum architecture search}.
\newblock \emph{\bibinfo{journal}{Machine Learning: Science and Technology}}
  \textbf{\bibinfo{volume}{2}}, \bibinfo{pages}{045027} (\bibinfo{year}{2021}).

\bibitem{ruiz2024quantum}
\bibinfo{author}{Ruiz, F. J.~R.} \emph{et~al.}
\newblock \bibinfo{title}{Quantum circuit optimization with alphatensor}
  (\bibinfo{year}{2024}).

\bibitem{NEURIPS2021_97244127}
\bibinfo{author}{Ostaszewski, M.}, \bibinfo{author}{Trenkwalder, L.~M.},
  \bibinfo{author}{Masarczyk, W.}, \bibinfo{author}{Scerri, E.} \&
  \bibinfo{author}{Dunjko, V.}
\newblock \bibinfo{editor}{Ranzato, M.}, \bibinfo{editor}{Beygelzimer, A.},
  \bibinfo{editor}{Dauphin, Y.}, \bibinfo{editor}{Liang, P.} \&
  \bibinfo{editor}{Vaughan, J.~W.} (eds) \emph{\bibinfo{title}{Reinforcement
  learning for optimization of variational quantum circuit architectures}}.
\newblock (eds \bibinfo{editor}{Ranzato, M.}, \bibinfo{editor}{Beygelzimer,
  A.}, \bibinfo{editor}{Dauphin, Y.}, \bibinfo{editor}{Liang, P.} \&
  \bibinfo{editor}{Vaughan, J.~W.}) \emph{\bibinfo{booktitle}{Advances in
  Neural Information Processing Systems}}, Vol.~\bibinfo{volume}{34},
  \bibinfo{pages}{18182--18194} (\bibinfo{publisher}{Curran Associates, Inc.},
  \bibinfo{year}{2021}).

\bibitem{altmann2024challenges}
\bibinfo{author}{Altmann, P.} \emph{et~al.}
\newblock \bibinfo{title}{Challenges for reinforcement learning in quantum
  circuit design} (\bibinfo{year}{2024}).

\bibitem{incudini2023automatic}
\bibinfo{author}{Incudini, M.} \emph{et~al.}
\newblock \bibinfo{title}{Automatic and effective discovery of quantum kernels}
  (\bibinfo{year}{2023}).

\bibitem{Schuld_2021}
\bibinfo{author}{Schuld, M.}, \bibinfo{author}{Sweke, R.} \&
  \bibinfo{author}{Meyer, J.~J.}
\newblock \bibinfo{title}{Effect of data encoding on the expressive power of
  variational quantum-machine-learning models}.
\newblock \emph{\bibinfo{journal}{Physical Review A}}
  \textbf{\bibinfo{volume}{103}} (\bibinfo{year}{2021}).

\bibitem{Schuld_sup_a_kernels}
\bibinfo{author}{Schuld, M.}
\newblock \bibinfo{title}{Supervised quantum machine learning models are kernel
  methods} (\bibinfo{year}{2021}).

\bibitem{PRXQuantum5020328}
\bibinfo{author}{Nguyen, Q.~T.} \emph{et~al.}
\newblock \bibinfo{title}{Theory for equivariant quantum neural networks}.
\newblock \emph{\bibinfo{journal}{PRX Quantum}} \textbf{\bibinfo{volume}{5}},
  \bibinfo{pages}{020328} (\bibinfo{year}{2024}).

\bibitem{sutton2018reinforcement}
\bibinfo{author}{Sutton, R.~S.} \& \bibinfo{author}{Barto, A.~G.}
\newblock \emph{\bibinfo{title}{Reinforcement learning: An introduction}}
  (\bibinfo{publisher}{MIT press}, \bibinfo{year}{2018}).

\bibitem{schulman2017proximal}
\bibinfo{author}{Schulman, J.}, \bibinfo{author}{Wolski, F.},
  \bibinfo{author}{Dhariwal, P.}, \bibinfo{author}{Radford, A.} \&
  \bibinfo{author}{Klimov, O.}
\newblock \bibinfo{title}{Proximal policy optimization algorithms}
  (\bibinfo{year}{2017}).

\bibitem{GARCIA1989335}
\bibinfo{author}{García, C.~E.}, \bibinfo{author}{Prett, D.~M.} \&
  \bibinfo{author}{Morari, M.}
\newblock \bibinfo{title}{Model predictive control: Theory and practice—a
  survey}.
\newblock \emph{\bibinfo{journal}{Automatica}} \textbf{\bibinfo{volume}{25}},
  \bibinfo{pages}{335--348} (\bibinfo{year}{1989}).

\bibitem{kreplin2023squlearn}
\bibinfo{author}{Kreplin, D.~A.}, \bibinfo{author}{Willmann, M.},
  \bibinfo{author}{Schnabel, J.}, \bibinfo{author}{Rapp, F.} \&
  \bibinfo{author}{Roth, M.}
\newblock \bibinfo{title}{squlearn --- a python library for quantum machine
  learning} (\bibinfo{year}{2023}).

\bibitem{Hubreg_}
\bibinfo{author}{Hubregtsen, T.} \emph{et~al.}
\newblock \bibinfo{title}{Training quantum embedding kernels on near-term
  quantum computers} (\bibinfo{year}{2021}).

\bibitem{peters2021machine}
\bibinfo{author}{Peters, E.} \emph{et~al.}
\newblock \bibinfo{title}{Machine learning of high dimensional data on a noisy
  quantum processor} (\bibinfo{year}{2021}).

\bibitem{Haug_2023}
\bibinfo{author}{Haug, T.}, \bibinfo{author}{Self, C.~N.} \&
  \bibinfo{author}{Kim, M.~S.}
\newblock \bibinfo{title}{Quantum machine learning of large datasets using
  randomized measurements}.
\newblock \emph{\bibinfo{journal}{Machine Learning: Science and Technology}}
  \textbf{\bibinfo{volume}{4}}, \bibinfo{pages}{015005} (\bibinfo{year}{2023}).

\bibitem{optuna_2019}
\bibinfo{author}{Akiba, T.}, \bibinfo{author}{Koyama, M.} \emph{et~al.}
\newblock \bibinfo{title}{Optuna: A next-generation hyperparameter optimization
  framework}.
\newblock \emph{\bibinfo{journal}{Proceedings of the 25th {ACM} {SIGKDD}
  International Conference on Knowledge Discovery and Data Mining}}
  (\bibinfo{year}{2019}).

\bibitem{kreplin2023reduction}
\bibinfo{author}{Kreplin, D.~A.} \& \bibinfo{author}{Roth, M.}
\newblock \bibinfo{title}{Reduction of finite sampling noise in quantum neural
  networks}.
\newblock \emph{\bibinfo{journal}{Quantum}} \textbf{\bibinfo{volume}{8}},
  \bibinfo{pages}{1385} (\bibinfo{year}{2024}).

\bibitem{bowles2024better}
\bibinfo{author}{Bowles, J.}, \bibinfo{author}{Ahmed, S.} \&
  \bibinfo{author}{Schuld, M.}
\newblock \bibinfo{title}{Better than classical? the subtle art of benchmarking
  quantum machine learning models} (\bibinfo{year}{2024}).

\bibitem{Glick2021hdc}
\bibinfo{author}{Glick, J.} \emph{et~al.}
\newblock \bibinfo{title}{Covariant quantum kernels for data with group
  structure}.
\newblock \emph{\bibinfo{journal}{Nature Physics}}
  \textbf{\bibinfo{volume}{20}}, \bibinfo{pages}{1--5} (\bibinfo{year}{2024}).

\bibitem{Chen_2016}
\bibinfo{author}{Chen, T.} \& \bibinfo{author}{Guestrin, C.}
\newblock \bibinfo{title}{Xgboost: A scalable tree boosting system}.
\newblock \emph{\bibinfo{journal}{Proceedings of the 22nd ACM SIGKDD
  International Conference on Knowledge Discovery and Data Mining}}
  (\bibinfo{year}{2016}).

\bibitem{tensorflow2015whitepaper}
\bibinfo{author}{Abadi, M.}, \bibinfo{author}{Zheng, X.} \emph{et~al.}
\newblock \bibinfo{title}{{TensorFlow}: Large-scale machine learning on
  heterogeneous systems} (\bibinfo{year}{2015}).
\newblock \bibinfo{note}{Software available from tensorflow.org}.

\bibitem{scikitlearn}
\bibinfo{author}{Pedregosa, F.}, \bibinfo{author}{Duchesnay, E.} \emph{et~al.}
\newblock \bibinfo{title}{Scikit-learn: Machine learning in {P}ython}.
\newblock \emph{\bibinfo{journal}{Journal of Machine Learning Research}}
  \textbf{\bibinfo{volume}{12}}, \bibinfo{pages}{2825--2830}
  (\bibinfo{year}{2011}).

\bibitem{xiao2017fashionmnist}
\bibinfo{author}{Xiao, H.}, \bibinfo{author}{Rasul, K.} \&
  \bibinfo{author}{Vollgraf, R.}
\newblock \bibinfo{title}{Fashion-mnist: a novel image dataset for benchmarking
  machine learning algorithms} (\bibinfo{year}{2017}).

\bibitem{Jerbi_2023}
\bibinfo{author}{Jerbi, S.} \emph{et~al.}
\newblock \bibinfo{title}{Quantum machine learning beyond kernel methods}.
\newblock \emph{\bibinfo{journal}{Nature Communications}}
  \textbf{\bibinfo{volume}{14}} (\bibinfo{year}{2023}).

\bibitem{QiskitCommunity2017}
\bibinfo{author}{{Qiskit Community}}.
\newblock \bibinfo{title}{Qiskit: {{An}} open-source framework for quantum
  computing} (\bibinfo{year}{2017}).
\newblock \urlprefix\url{https://github.com/Qiskit/qiskit}.

\end{thebibliography}

\begin{appendices}

\section{Reference circuits}
\label{app:ref_circuits}
\begin{figure*}[t]
    \centering
    \begin{minipage}{0.65\textwidth}
        \centering
        \includegraphics[width=\linewidth]{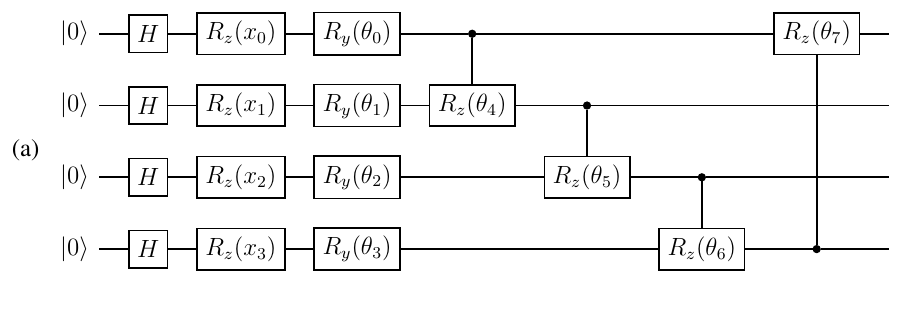}
    \end{minipage}
    \hfill
    \begin{minipage}{0.65\textwidth}
        \centering
        \includegraphics[width=\linewidth]{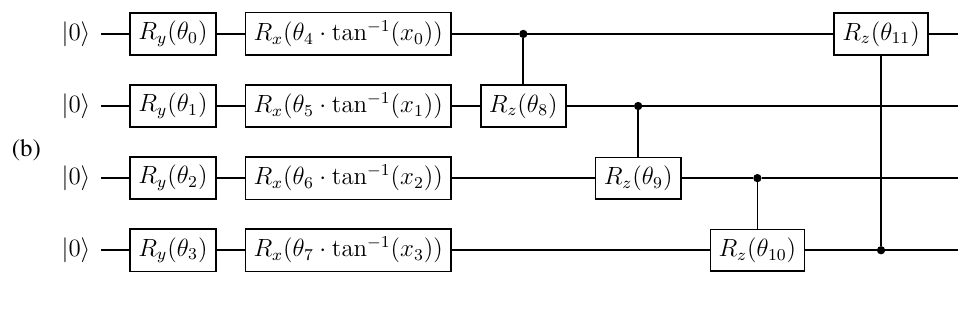}
    \end{minipage}
    
    \vskip\baselineskip
    
    \begin{minipage}{0.65\textwidth}
        \centering
        \includegraphics[width=\linewidth]{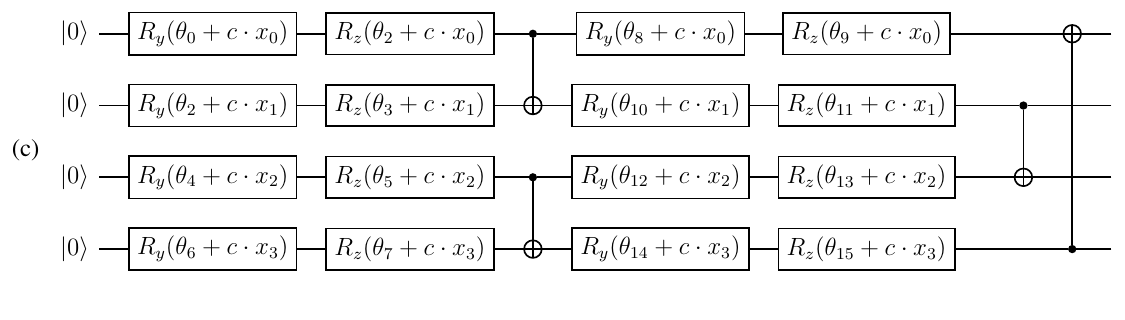}
    \end{minipage}
    \hfill
    \begin{minipage}{0.65\textwidth}
        \centering
        \includegraphics[width=\linewidth]{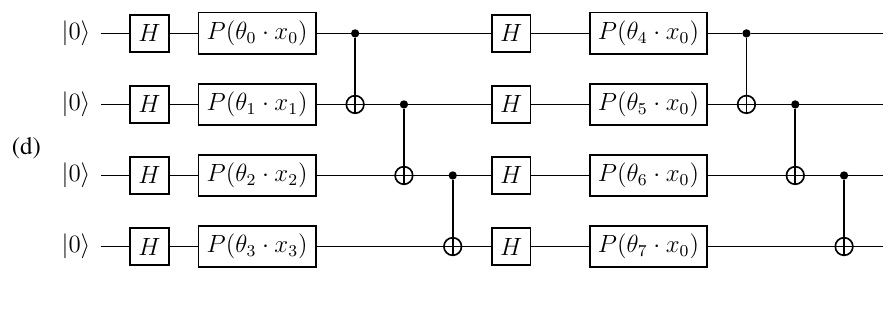}
    \end{minipage}
    
    \caption{Reference circuit architectures from the literature. (a) is taken from~\cite{Hubreg_}, (b) from~\cite{kreplin2023reduction}, (c) from~\cite{Haug_2023}, and (d) from~\cite{peters2021machine}. All graphics show one layer-block of the respective architectures.}
    \label{fig: ref circuits}
\end{figure*}

Figure~\ref{fig: ref circuits} shows the different reference circuits from the literature used in this work. All layouts show one block of the respective architectures, which can be repeated several times. In this work, all reference circuits used in the experiments in Sec.~\ref{sec: numerical results} consist of two blocks for each architecture. 
The fifth encoding circuit architecture is the \href{https://docs.quantum.ibm.com/api/qiskit/qiskit.circuit.library.ZFeatureMap}{ZFeatureMap} taken from Qiskit~\cite{QiskitCommunity2017}.

These reference circuits incorporate popular design choices. For example, they use linear nearest-neighbor entangling and parametrized data encoding, with the choice between linear or non-linear encoding depending on the circuit. Alternatively, they may use a fixed encoding with parametrized scaling.

We introduced a slight modification to the circuit shown in Fig.~\ref{fig: ref circuits}(b), in comparison to the original circuit used in~\cite{kreplin2023reduction}. Specifically, we used $\arctan(x)$ instead of $\arccos(x)$ as non-linear encoding strategy to align with the action set proposed for the RL algorithm.

\end{appendices}

\end{document}